\documentclass[twocolumn,nofootinbib]{aastex62}
\usepackage{aas_macros}
\usepackage{graphicx}
\usepackage{dcolumn}
\usepackage{bm}
\usepackage{amsmath}
\usepackage{float}
\usepackage{lipsum}
\usepackage{color}
\usepackage{times}
\usepackage{textcomp}
\usepackage{latexsym}

\newcommand{\approptoinn}[2]{\mathrel{\vcenter{
  \offinterlineskip\halign{\hfil$##$\cr
    #1\propto\cr\noalign{\kern2pt}#1\sim\cr\noalign{\kern-2pt}}}}}

\newcommand{\FAP}{\text{FAP}}
\newcommand{\ra}{\alpha}
\newcommand{\dec}{\delta}

\newcommand{\vtheta}{{\bm \theta}}
\newcommand{\vbeta}{{\bm \beta}}
\newcommand{\der}{\mathrm{d}}
\newcommand{\RLU}{\mathcal{R}^L_U}

\begin{document}

\title{Beyond the detector horizon: Forecasting gravitational-wave strong lensing}
\date{\today}

\author{A.~Renske~A.~C.~Wierda}
\email{a.r.a.c.wierda@uu.nl}
\affiliation{Institute for Gravitational and Subatomic Physics (GRASP), Department of Physics, Utrecht University, Princetonplein 1, 3584 CC Utrecht, The Netherlands} 
\author{Ewoud~Wempe}
\email{wempe@astro.rug.nl}
\affiliation{Kapteyn Astronomical Institute, University of Groningen, P.O Box 800, 9700 AV Groningen, The Netherlands} 
\author{Otto~A.~Hannuksela}
\email{o.hannuksela@nikhef.nl}
\affiliation{Department of Physics, The Chinese University of Hong Kong, Shatin, NT, Hong KongL}
\affiliation{Nikhef – National Institute for Subatomic Physics, Science Park, 1098 XG Amsterdam, The Netherlands}
\affiliation{Institute for Gravitational and Subatomic Physics (GRASP), Department of Physics, Utrecht University, Princetonplein 1, 3584 CC Utrecht, The Netherlands} 
\author{Léon~V.~E.~Koopmans}
\email{l.v.e.koopmans@rug.nl}
\affiliation{Kapteyn Astronomical Institute, University of Groningen, P.O Box 800, 9700 AV Groningen, The Netherlands} 
\author{Chris~Van~Den~Broeck}
\email{c.f.f.vandenbroeck@uu.nl}
\affiliation{Nikhef – National Institute for Subatomic Physics, Science Park, 1098 XG Amsterdam, The Netherlands}
\affiliation{Institute for Gravitational and Subatomic Physics (GRASP), Department of Physics, Utrecht University, Princetonplein 1, 3584 CC Utrecht, The Netherlands}

\begin{abstract}
\noindent
When gravitational waves pass near massive astrophysical objects, they can be gravitationally lensed. The lensing can split them into multiple wave-fronts, magnify them, or imprint beating patterns on the waves. Here we focus on the multiple images produced by strong lensing. In particular, we investigate strong lensing forecasts, the rate of lensing, and the role of lensing statistics in strong lensing searches. 
Overall, we find a reasonable rate of lensed detections for double, triple, and quadruple images at the LIGO--Virgo--KAGRA design sensitivity. 
We also report the rates for A+ and LIGO Voyager and briefly comment on potential improvements due to the inclusion of sub-threshold triggers. 
We find that most galaxy-lensed events originate from redshifts $z \sim 1-4$ and report the expected distribution of lensing parameters for the observed events. Besides forecasts, we investigate the role of lensing forecasts in strong lensing searches, which explore repeated event pairs. One problem associated with the searches is the rising number of event pairs, which leads to a rapidly increasing false alarm probability. 
We show how knowledge of the expected galaxy lensing time delays in our searches allow us to tackle this problem. 
Once the time delays are included, the false alarm probability increases linearly (similar to non-lensed searches) instead of quadratically with time, significantly improving the search.
For galaxy cluster lenses, the improvement is less significant. 
The main uncertainty associated with these forecasts are the merger-rate density estimates at high redshift, which may be better resolved in the future.
\end{abstract}

\section{Introduction}

Similarly to light, gravitational waves can be gravitationally lensed by massive astrophysical objects, e.g., galaxies and galaxy clusters~\citep{Ohanian:1974ys,Thorne:1982cv,Deguchi:1986zz,Wang:1996as,Nakamura:1997sw,Takahashi:2003ix}.
Lensing changes the gravitational-wave amplitude without changing its frequency evolution~\citep{Deguchi:1986zz,Wang:1996as,Nakamura:1997sw,Takahashi:2003ix, Dai:2017huk,Ezquiaga:2020gdt}.
Moreover, strong lensing produces multiple images observable at the detectors as repeated events separated by minutes to months when lensed by galaxies~\citep{Ng:2017yiu,Li:2018prc,Oguri:2018muv}, and up to years
when lensed by galaxy clusters~\citep{Smith:2017mqu,Smith:2018gle,Smith:2019dis,Robertson:2020mfh,Ryczanowski:2020mlt}.

While much of the gravitational-wave lensing theory is similar to electromagnetic lensing, the detection methodologies and the science case are different. For example, in light lensing, one can observe strong lensing by discerning multiple images with telescope imaging. In GW lensing, we observe strongly lensed GWs as repeated events that can be identified with GW templates inaccessible to electromagnetic searches~\citep{Haris:2018vmn,Hannuksela:2019kle,Dai:2020tpj,Liu:2020par,Lo:2021nae,Janquart:2021qov}. The principal methodologies to detect gravitational-wave lensing with ground-based detectors have been developed in recent years~\citep{Cao:2014oaa,Lai:2018rto,Haris:2018vmn,Hannuksela:2019kle,Pang:2020qow, Pagano:2020rwj, Hannuksela:2020xor,Dai:2020tpj,Liu:2020par,Lo:2021nae,Janquart:2021qov}. Moreover, the LIGO-Vigro Collaboration (LVC) performed the first comprehensive search for gravitational-wave lensing signatures in the first half of the third LIGO-Virgo observing run recently~\citep{Abbott:2021iab}. 

If detected, gravitational-wave lensing may enable several exciting scientific frontiers such as localisation of merging black holes to sub-arcsecond precision~\citep{Hannuksela:2020xor}, precision cosmography studies~\citep{Sereno:2011ty,Liao:2017ioi,Cao:2019kgn,Li:2019rns,Hannuksela:2020xor}, precise tests of the speed of gravitational-wave propagation~\citep{Baker:2016reh,Fan:2016swi,Mukherjee:2019wfw,Mukherjee:2019wcg}, tests of the gravitational-wave polarization content~\citep{Goyal:2020bkm}, and detecting intermediate-mass or primordial black holes~\citep{Lai:2018rto,Diego:2019rzc,Oguri:2020ldf}. They may also be useful in lens modelling by allowing one to break the mass-sheet degeneracy~\citep{Cremonese:2021puh}. 

Recent strongly lensed gravitational-wave forecasts have predicted gravitational-wave lensing at a reasonable rate at design sensitivity of the Advanced LIGO and Advanced Virgo detectors~\citep{Ng:2017yiu,Li:2018prc,Oguri:2018muv,Xu:2021bfn,Mukherjee:2021} 
(see also~\citet{Smith:2017mqu,Smith:2018gle,Smith:2019dis,Robertson:2020mfh,Ryczanowski:2020mlt} for estimates for galaxy clusters). 
In addition, \citet{Xu:2021bfn} studied lensing forecasts in the context of probing the black hole and lens populations, while \citet{Mukherjee:2021} studied the impact of the binary coalescence times on the rate of lensing. 
\citet{Haris:2018vmn} characterized the distribution of lensed events. 
Here we further investigate strong lensing forecasts with a focus on the lensing science case and searches. 

The science targets depend on the number of identified pairs. Suppose we have access to four lensed images of a gravitational-wave event. In that case, we might localise the gravitational-wave event to its host galaxy by comparing the image properties of the lensed wave with those produced by galaxies independently observed in the electromagnetic bands~\citep{Hannuksela:2020xor}. Two images might still allow us to constrain the number of candidates~\citep{Sereno:2011ty,Yu:2020agu}, but to a lesser degree as we will need to rely mainly on the magnification ratios to pinpoint the source location.\footnote{A search for a system lensed by a galaxy cluster might also be promising, even with two images~\citep{Smith:2017mqu,Smith:2018gle,Smith:2019dis,Robertson:2020mfh,Ryczanowski:2020mlt}.} More images also allow for better cosmography~\citep{Sereno:2011ty,Liao:2017ioi,Cao:2019kgn,Li:2019rns,Hannuksela:2020xor} and polarization tests~\citep{Goyal:2020bkm}.  

Therefore, in Sec.~\ref{sec:lensed_rates}, we investigate the number of images discoverable in LIGO~\citep{Harry:2010zz,TheLIGOScientific:2014jea,TheVirgo:2014hva,Martynov:2016fzi,TheLIGOScientific:2016agk}, Virgo~\citep{TheVirgo:2014hva}, KAGRA~\citep{Somiya:2011np,Aso:2013eba,Akutsu:2020his}, A+~\citep{Abbott:2020qfu}, and LIGO Voyager~\citep{Adhikari:2020gft}. 
In particular, we might identify two or more super-threshold triggers when we search for multiply imaged, strongly lensed gravitational waves~\citep{Li:2018prc}. However, it is also entirely plausible to observe some of these multiple images below the usual noise threshold as sub-threshold triggers~\citep{Li:2019osa,McIsaac:2019use,Mukherjee:2021}.
Thus, we also comment on sub-threshold triggers.

Another important question to address is how lensing forecasts can help the strong lensing parameter estimation (see~\citet{Haris:2018vmn,Hannuksela:2019kle,Liu:2020par,Lo:2021nae,Janquart:2021qov}). In particular, unlensed events can mimic a strongly lensed event by chance, resulting in a false alarm. The probability of a false alarm increases as we detect more events ($\propto N^2$, number of events squared) until the likelihood of a false alarm occurring becomes inevitable. However, we show how incorporating knowledge of the galaxy lensing time delay can significantly improve searches so that the false alarm increases at the same rate as it does for usual searches (Sec.~\ref{sec:FAP}). 

While the time delay effect has been investigated, e.g., in \citet{Haris:2018vmn}, it has usually been discussed in the context of an additional improvement upon the usual searches. 
Here we point out how, without the information of the lensing time-delay distribution, strong lensing searches may rapidly become intractable due to the growing number of candidate pairs. 

Finally, we report the redshift distribution of lensed events and the Einstein radii of the systems that lens them and briefly comment on the science case (Sec.~\ref{sec:redshift}). 
We conclude in Sec.~\ref{sec:conclusions}.
Throughout this paper, we assume a flat $\Lambda$CDM cosmology with $H_0=70 \, \rm km \, s^{-1} \, Mpc^{-1}$ and $\Omega_m = 0.31$, and all uncertainties quoted are at the 90 \% confidence level.

\section{Catalogue of lensed events} 
\label{sec:lensed_catalogue}

We model the mass distribution of binary black holes following the observational results for the \textsc{Power Law + Peak model} of~\citet{GWTC2:rates}, setting the mass power-law index $\alpha=2.63$, mass ratio power-law index $\beta_q = 1.26$, low-mass tapering at $\delta_m = 4.82$~$\rm M_\odot$, minimum and maximum masses $m_{\rm min}=4.59$~$\rm M_\odot$ and $m_{\rm max}=86.22$~$\rm M_\odot$, and a Gaussian peak at $\mu_{\rm m}=33.07$~$\rm M_\odot$ with a width $\sigma_m=5.69$~$\rm M_\odot$, for a fraction of the population $\lambda_{\rm peak} = 0.10$. These values are consistent with the LIGO--Virgo population studies~\citep{GWTC2:rates}. We adopt a fit to the Population I/II star merger-rate density normalized to the local merger-rate density following~\citet{Oguri:2018muv}, 
\begin{equation}
    \label{eqn:mergerrate}
    \mathcal{R}_\textrm{m}(z_s) = \frac{\mathcal{R}_0(b_4+1)e^{b_2 z_s}}{b_4+e^{b_3 z_s}}  {\rm Gpc^{-3}}\, {\rm yr^{-1}} \,,
\end{equation}
where $\mathcal{R}_0 $ is the local merger-rate density, and $b_2=1.6$, $b_3=2.1$, and $b_4=30$ are fitting parameters. We take the local merger-rate density to be consistent with the local merger-rate observations $\mathcal{R}_0 = 23.9^{+14.3}_{-8.6} \, \rm Gpc^{-3} \, yr^{-1}$~\citep{GWTC2:rates}, where we take the uncertainty to be constant with redshift. 

The galaxy lens population follows the SDSS galaxy catalogue~\citep{Collett:2015roa}, and we loosely follow \citet{Haris:2018vmn} in the derivation of the lens population and our sampling procedure. 
The strong lensing optical depth~\citep{Haris:2018vmn}
\begin{equation}
    \label{eqn:opticaldepth}
    \tau(z_s) = 4.17 \times 10^{-6} \left ( \frac{D_\textrm{c}(z_s)}{\rm Gpc} \right )^3\,,
\end{equation}
where $D_\textrm{c}(z_s)$ is the comoving distance\footnote{Note that the optical depth definition here refers to the probability that a given event is lensed irrespective of whether it is detected; the information about the binary black hole population and the selection bias is included separately in the rate computations (Appendix~\ref{app:lensed_rate_derivation}).}. 
Note that here we have approximated the optical depth using the singular isothermal sphere (SIS) lens model; including ellipticity may yield a $\sim 5-10\, \%$ correction~\citep{More:2011rn, Xu:2021bfn}. 

To facilitate quadruply imaged sources (lensed events that are split into four images) and realistic lens models, we adopt a power-law ellipsoidal mass distribution with external shear to approximate our lensing galaxies, available in~\textsc{lenstronomy}~\citep{Birrer:2018xgm}. 
Specifically, we assume SDSS velocity dispersion and axis ratio profiles of elliptical galaxies in the local Universe~\citep{Collett:2015roa}, a 0.05 spread (one standard deviation) on the measurement of each shear component, and a typical power-law density slope with a mean slope $\gamma=2$ with $0.2$ spread~\citep{Koopmans:2009} (see Appendices \ref{app:bbh} \& \ref{app:lenses}, for the full population details). 

To model the rate of \emph{detectable} events, we employ Monte Carlo importance sampling to sample the binary and the lens population (see Appendix \ref{app:lensed_rate_derivation} for the full details), selecting only events that pass the detection threshold on the signal-to-noise ratio (SNR). We assume spin-less binary black holes and adopt the \textsc{IMRPhenomD}~\citep{Husa:2015iqa, Khan:2015jqa} waveform model.  
Our results partially extend previous forecast studies~\citep[e.g.,][]{Haris:2018vmn,Li:2018prc}, by considering an updated mass-population model, a network of detectors, and a power-law ellipsoidal mass distribution with external shear. 

\section{Lensed rates}
\label{sec:lensed_rates}
\begin{table*}[t]
    \centering
    \begin{tabular}{r r l l l l l}
        \hline \noalign{\smallskip}
        \multicolumn{2}{r}{Observed rates} & L & L/H  & L/H/V/K & L/H/V/K (A+) & L/H/V/K (Voyager) \\
        \noalign{\smallskip}\hline \hline \noalign{\smallskip}
        Lensed events: & total & $0.21^{+0.10}_{-0.07}$ $\text{yr}^{-1}$ & $0.65^{+0.32}_{-0.22}$ $\text{yr}^{-1}$ & $1.3^{+0.6}_{-0.4}$ $\text{yr}^{-1}$& $3.3^{+1.7}_{-1.1}$ $\text{yr}^{-1}$ & $16.8^{+8.4}_{-5.6}$ $\text{yr}^{-1}$ \\[2 pt]
        \multicolumn{2}{r}{double} & $0.17^{+0.08}_{-0.06}$ $\text{yr}^{-1}$ & $0.50^{+0.25}_{-0.17}$ $\text{yr}^{-1}$ & $0.92^{+0.46}_{-0.31}$ $\text{yr}^{-1}$ & $2.5^{+1.2}_{-0.8}$ $\text{yr}^{-1}$ & $13.1^{+6.5}_{-4.4}$ $\text{yr}^{-1}$ \\[2 pt]
        \multicolumn{2}{r}{triple} & $0.032^{+0.016}_{-0.011}$ $\text{yr}^{-1}$ & $0.11^{+0.06}_{-0.04}$ $\text{yr}^{-1}$ & $0.23^{+0.12}_{-0.08}$ $\text{yr}^{-1}$ & $0.55^{+0.28}_{-0.19}$ $\text{yr}^{-1}$ & $2.0^{+1.0}_{-0.7}$ $\text{yr}^{-1}$ \\[2 pt]
        \multicolumn{2}{r}{quadruple} & $0.011^{+0.005}_{-0.004}$ $\text{yr}^{-1}$ & $0.038^{+0.019}_{-0.013}$ $\text{yr}^{-1}$ & $0.12^{+0.06}_{-0.04}$ $\text{yr}^{-1}$ & $0.30^{+0.15}_{-0.10}$ $\text{yr}^{-1}$ & $1.6^{+0.8}_{-0.6}$ $\text{yr}^{-1}$\\
        \noalign{\smallskip} \hline \noalign{\smallskip}
        \multicolumn{2}{r}{Unlensed events} & $370$ $\text{yr}^{-1}$ & $1.1 \times 10^3$ $\text{yr}^{-1}$ & $1.9 \times 10^3$ $\text{yr}^{-1}$ & $5.8 \times 10^3$ $\text{yr}^{-1}$ & $31 \times 10^3$ $\text{yr}^{-1}$ \\[2 pt]
        \multicolumn{2}{r}{Relative occurrence} & 1 : 1760 & 1 : 1650 & 1 : 1500 & 1 : 1740 & 1 : 1830 \\
        \noalign{\smallskip} \hline
    \end{tabular}
    \caption{
    The observed lensed event rates for different detector networks and detector sensitivities (LIGO Livingston [L] and Hanford [H] at their design, A+, and Voyager sensitivities; Virgo [V] and KAGRA [K] are always at their design sensitivities), categorised according to the \textit{observed} number of super-threshold images. 
    All uncertainties are at the 90 \% confidence level and are a direct consequence of the uncertainty in the local merger-rate density.
    Unless otherwise specified, design sensitivity is assumed. 
    The rates can be subject to some uncertainties introduced by different merger-rate density models, the choice of the detection threshold, and detector down-time. 
    However, we also report the relative rate of occurrences, which we expect to be subject to less uncertainty. 
    }
    \label{tab:rates}
\end{table*}

\begin{table*}[t]
    \centering
    \begin{tabular}{r r l l l l l}
        \hline \noalign{\smallskip}
        \multicolumn{2}{r}{Observed rates} & L & L/H & L/H/V/K & L/H/V/K (A+) & L/H/V/K (Voyager) \\
        \noalign{\smallskip}\hline \hline \noalign{\smallskip}
        Lensed events: & total & $0.30^{+0.15}_{-0.10}$ $\text{yr}^{-1}$ & $0.90^{+0.45}_{-0.30}$ $\text{yr}^{-1}$ & $1.7^{+0.9}_{-0.6}$ $\text{yr}^{-1}$& $4.3^{+2.1}_{-1.5}$ $\text{yr}^{-1}$ & $19.9^{+9.9}_{-6.7}$ $\text{yr}^{-1}$ \\[2 pt]
        \multicolumn{2}{r}{double} & $0.23^{+0.12}_{-0.08}$ $\text{yr}^{-1}$ & $0.67^{+0.33}_{-0.22}$ $\text{yr}^{-1}$& $1.2^{+0.6}_{-0.4}$ $\text{yr}^{-1}$ & $3.2^{+1.6}_{-1.1}$ $\text{yr}^{-1}$ & $15.6^{+7.8}_{-5.2}$ $\text{yr}^{-1}$ \\[2 pt]
        \multicolumn{2}{r}{triple} & $0.054^{+0.027}_{-0.018}$ $\text{yr}^{-1}$ & $0.17^{+0.08}_{-0.06}$ $\text{yr}^{-1}$& $0.32^{+0.16}_{-0.11}$ $\text{yr}^{-1}$ & $0.71^{+0.35}_{-0.24}$ $\text{yr}^{-1}$ & $2.3^{+1.1}_{-0.8}$ $\text{yr}^{-1}$ \\[2 pt]
        \multicolumn{2}{r}{quadruple} & $0.015^{+0.008}_{-0.005}$ $\text{yr}^{-1}$ & $0.061^{+0.031}_{-0.021}$ $\text{yr}^{-1}$ & $0.18^{+0.09}_{-0.06}$ $\text{yr}^{-1}$ & $0.43^{+0.21}_{-0.14}$ $\text{yr}^{-1}$ & $2.0^{+1.0}_{-0.7}$ $\text{yr}^{-1}$\\
        \noalign{\smallskip} \hline \noalign{\smallskip}
        \multicolumn{2}{r}{Relative occurrence} & 1 : 1210 & 1 : 1180 & 1 : 1100 & 1 : 1350 & 1 : 1540 \\[2 pt]
        \multicolumn{2}{r}{Overall increase} & 45 \% & 39 \% & 36 \% & 29 \% & 19\% \\
        \noalign{\smallskip} \hline
    \end{tabular}
    \caption{
    The observed lensed event rates for different detector networks and detector sensitivities (LIGO Livingston [L] and Hanford [H] at their design, A+, and Voyager sensitivities; Virgo [V] and KAGRA [K] are always at their design sensitivities), including events with an SNR $>7$. 
    Note that here we presume that SNR $>7$ is an indicative proxy for a detection using sub-threshold searches~\citep{Li:2019osa,McIsaac:2019use,Abbott:2021iab}, but a more comprehensive study inspecting selection and estimates based on the false alarm probability will be required to quantify the precise improvement. 
    Unless otherwise specified, design sensitivity is assumed.
    }
    \label{tab:subrates}
\end{table*}

We classify a super-threshold event as an event trigger with a network SNR $\geq 8$ (for a discussion on the suitability of this SNR limit, see, e.g.,~\citet{Abbott:2020qfu}). 
Assuming the two LIGO, the Virgo, and the KAGRA detectors operating 100 \% of the time at design sensitivity, we find that the total \emph{observed} rate of lensed events is $1.3^{+0.6}_{-0.4}$~$\rm yr^{-1}$. The observed rate of unlensed events is $\sim 1900$ $\rm yr^{-1}$, which gives us a relative rate of 1 lensed event for every 1500 unlensed event detections. 
The relative rate of lensed-to-unlensed detections is broadly consistent with findings from, e.g.,~\citet{Li:2018prc, Oguri:2018muv}. 
The expected event rates for variable numbers of super-threshold images and different sensitivities (design, the A+ detector upgrade, and the planned LIGO-Voyager detector) are given in Table~\ref{tab:rates}.
The uncertainties in the observed rate here are a direct consequence of the uncertainty in the local merger-rate density. 
Note that the rate of observed events here is increased by the network of detectors; the single-detector (LIGO Livingston) estimate for the rate of unlensed events is around $\sim 370$ events per year (consistent with, e.g.,~\citet{Xu:2021bfn}). 

Note that once detector down-time is included, the observed rate can drop by a factor of two or more. 
Moreover, there is some uncertainty in the choice of the detection threshold, in the sense that the usual templated searches classify the detection threshold based on the false alarm rate, and not the SNR~\cite[e.g.,][]{Abbott:2020qfu}. 
We expect that such uncertainties can shift the total observed rates by perhaps an additional factor of a few. 
However, the results can be re-scaled based on the fractional rate of lensed to unlensed events, which we expect to be less sensitive to detector down-time, the precise detection threshold, or the local merger-rate density. 

\begin{figure}[!b]
    \centering
    \includegraphics[width=\linewidth]{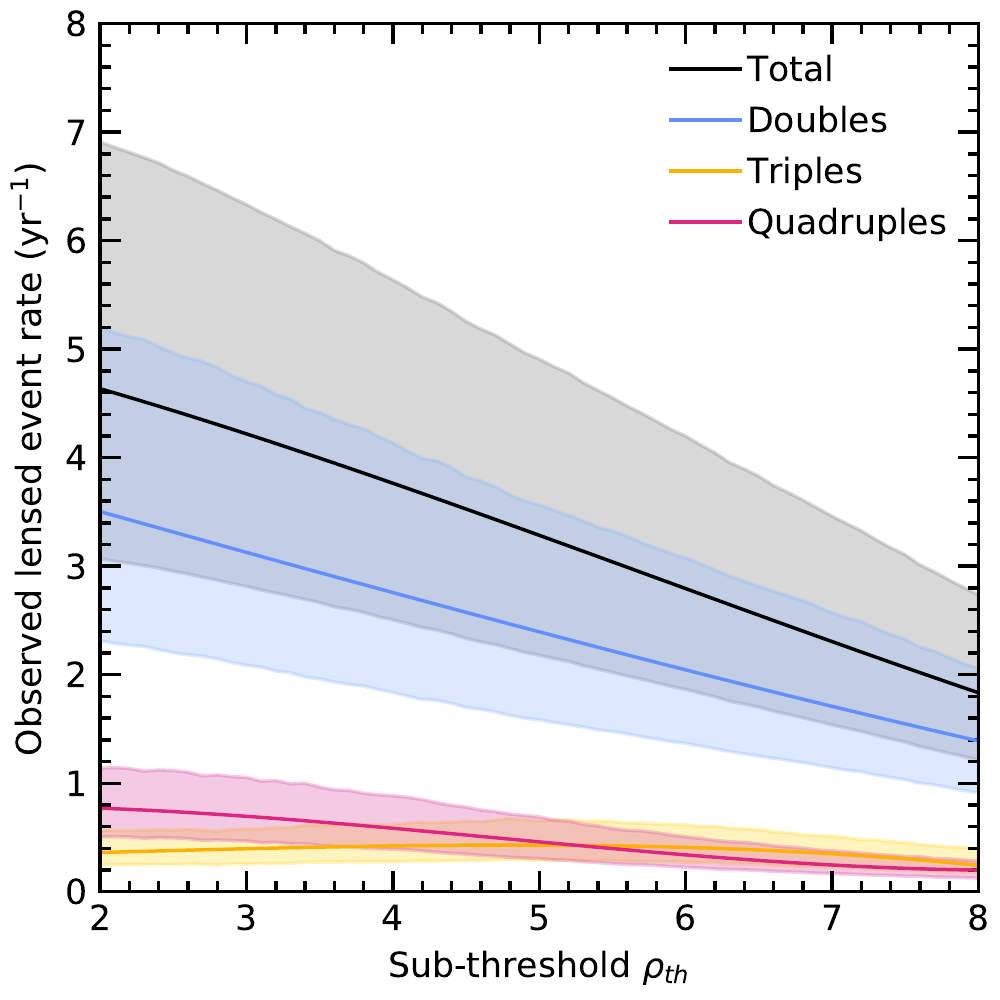}
    \caption{The observed lensed event rate as a function of detection threshold SNR $\rho_{th}$ for double (blue), triple (yellow) and quadruple (magenta) lensed event detections and the total rate (black). The observed rates, most notably the quadruple image detection rates, increase by several factors as the threshold SNR decreases.
    }
    \label{fig:supervssub}
\end{figure}

Targeted lensed searches, when at least one super-threshold counterpart image is available, may allow one to uncover so-called sub-threshold triggers below the usual noise threshold by reducing the background noise and glitch contribution~\citep{Li:2019osa, McIsaac:2019use}. We classify a sub-threshold event as an event trigger observed below a network SNR of 8, but above a network SNR of $\rho_{th}$, when at least one counter image with SNR $> 8$ is present. Since $\rho \propto d_L^{-1}$, \citet{Li:2019osa} provides an indicative increase in the effective distance of $\sim 15\%$ corresponding to $\rho_{th} = 7$. The expected event rates for variable numbers of detected images and detector sensitivities are given in Table~\ref{tab:subrates}. We find that the total number of observed quadruply lensed events, increases from $0.12^{+0.06}_{-0.04}$ $\text{yr}^{-1}$ to $0.18^{+0.09}_{-0.06}$ $\text{yr}^{-1}$, an increase of $51 \%$, when considering sub-threshold triggers. Furthermore, the total number of observed triply lensed events increases with $40 \%$ from $0.23^{+0.12}_{-0.08}$ $\text{yr}^{-1}$ to $0.32^{+0.16}_{-0.11}$ $\text{yr}^{-1}$ and for doubly lensed events there is an increase of $33 \%$ from $0.92^{+0.46}_{-0.31}$ $\text{yr}^{-1}$ to $1.2^{+0.6}_{-0.4}$ $\text{yr}^{-1}$. The ``double", ``triple" and ``quadruple" nomenclatures refer to the number of detected images, and not the number of images produced by the lens. The increase in detectable images further motivates follow-up sub-threshold searches~\citep{Li:2019osa,McIsaac:2019use}.

However, because the sub-threshold searches vary in their sensitivity and further improvements may still be possible, the SNR threshold choice may vary. Thus, a threshold of SNR $>7$ is not a flawless proxy for detection. 
For this reason, we also show the detectable rates for variable SNR thresholds (Fig.~\ref{fig:supervssub}).

\begin{figure*}[!th]
    \centering
    \includegraphics[width = \linewidth]{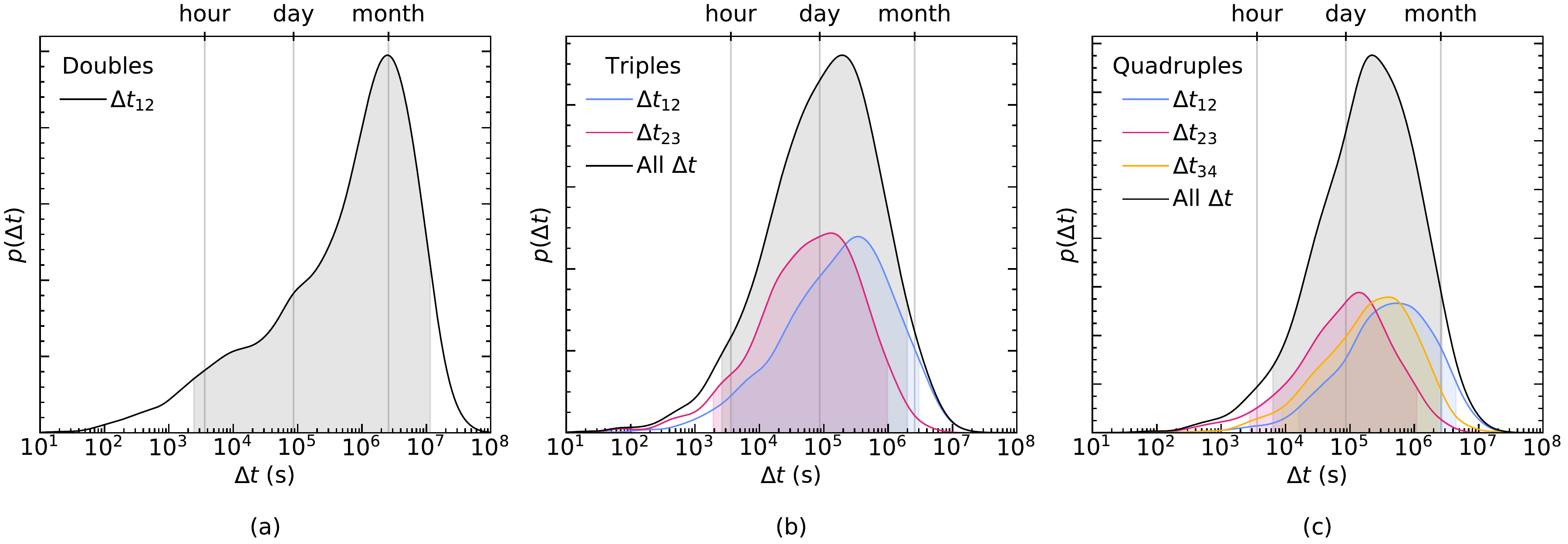}
    \caption{\textbf{(a)} The time-delay distribution for observed double images. The shaded regions give the 90 \% confidence intervals for $\Delta t_{12} \sim 1.5 \text{ hr} - 133 \text{ days}$. \textbf{(b)} The time-delay distributions (with confidence intervals) for observed triply lensed sources between the first two images (blue), between the second and third images (magenta) and the sum of those two (black). \textbf{(c)} The time-delay distributions (with confidence intervals) for observed quadruply lensed sources between the first two images (blue; $\sim 4.8 \text{ hr} - 52 \text{ days}$), the second and the third images (magenta; $\sim 0.8 \text{ hr} - 12 \text{ days}$), the last two images (yellow; $\sim 1.7 \text{ hr} - 30 \text{ days}$), and the total of the three (black; $\sim 2.3 \text{ hr} - 28 \text{ days}$). Generally, the time delay between the lensed pairs is $\lesssim 93 \, {\rm days}$. Knowledge of the time-delay distribution is particularly useful in improving strong lensing parameter estimation.
    } 
    \label{fig:time_delays}
\end{figure*}

\begin{figure}[!b]
    \centering
    \includegraphics[width = 0.95\linewidth]{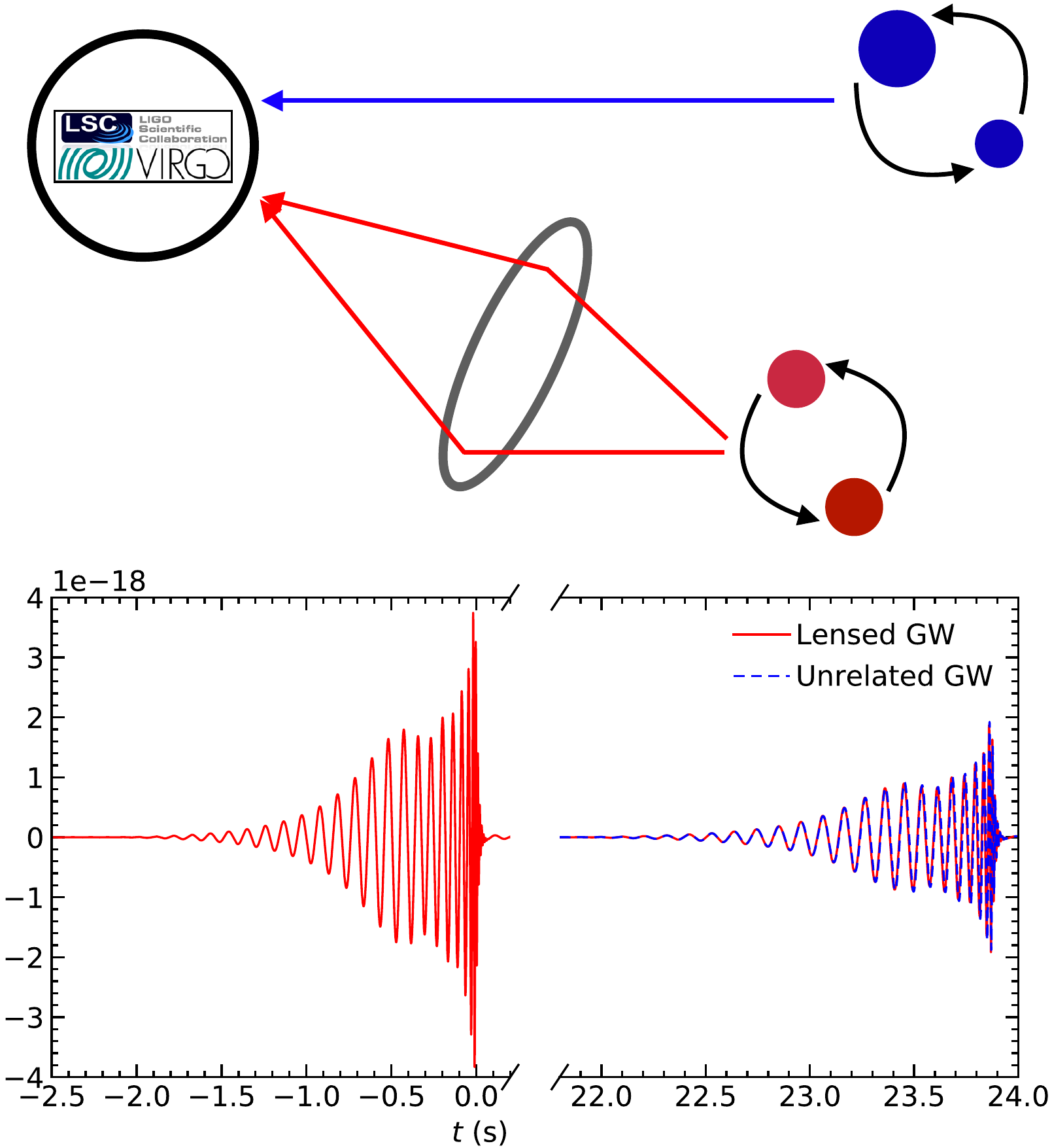}
    \caption{\emph{Graphical illustration of an unlensed event mimicking a strongly lensed event:} Two images of a single gravitationally lensed gravitational-wave event (red) and an unlensed event (blue). In this example, the unlensed gravitational wave signal (blue) is indistinguishable from the strongly lensed gravitational-wave images (right bottom panel). 
    Indeed, unlensed events can, in principle, resemble strongly lensed events, giving rise to strong lensing mimickers or "false alarms." 
    }
    \label{fig:waveforms}
\end{figure}

We note that the rate estimates are subject to further uncertainties due to a largely (observationally) unconstrained high-redshift merger-rate density. The merger-rate density can be modeled, for example, by presuming that the observed binary black hole population originates from Population-I/II stars, as we have done here. 
Still, there are variations to the specific predictions in the different models and population-synthesis simulations~\cite[e.g.,][]{Eldridge:2018nop,Neijssel:2019irh,Boco:2019teq,Santoliquido:2020axb,Abbott:2021iab,Mukherjee:2021}. Here we postpone the investigation of different model predictions and instead note that the rate of lensing will be constrained by direct observations of gravitational-wave lensing~\citep{Mukherjee:2021}, 
and to a degree by the stochastic gravitational-wave background~\citep{Buscicchio:2020bdq,Mukherjee:2020tvr,Buscicchio:2020cij,Abbott:2021iab}. This work focuses on the science case for gravitational-wave lensing, the strong lensing searches, and the relative improvement in the multiple-image detections due to detector upgrades.

\section{The lensing time-delay distribution and its effect on strong lensing searches}
\label{sec:FAP}

The expected observed time-delay distribution is a direct output of our mock catalogue of lensed events (Fig.~\ref{fig:time_delays}). 
To test whether two gravitational-wave events are lensed, one must show that the waves are identical within detector accuracy (save for an overall difference in the complex phase, arrival time, and amplitude), as expected of the lensing hypothesis~\citep{Haris:2018vmn,Hannuksela:2019kle,Dai:2020tpj,Liu:2020par,Lo:2021nae,Janquart:2021qov,Abbott:2021iab}. 
However, it is also possible for two waveforms to be near-identical within detector accuracy by chance, giving rise to strong lensing "mimickers" (see Fig.~\ref{fig:waveforms}, for an illustration). 
Here we demonstrate how the galaxy-lensing time-delay prior allows us to keep the strong lensing searches tractable. 

The time-delay distribution of the unlensed events follows a Poissonian process~\citep{Haris:2018vmn}. The distributions for lensed events are an output of our simulation (see Fig. \ref{fig:time_delays}).
Given an expected lensing time-delay distribution, this allows us to calculate a ranking statistic 
\begin{figure}[!th]
    \centering
    \includegraphics[width = \linewidth]{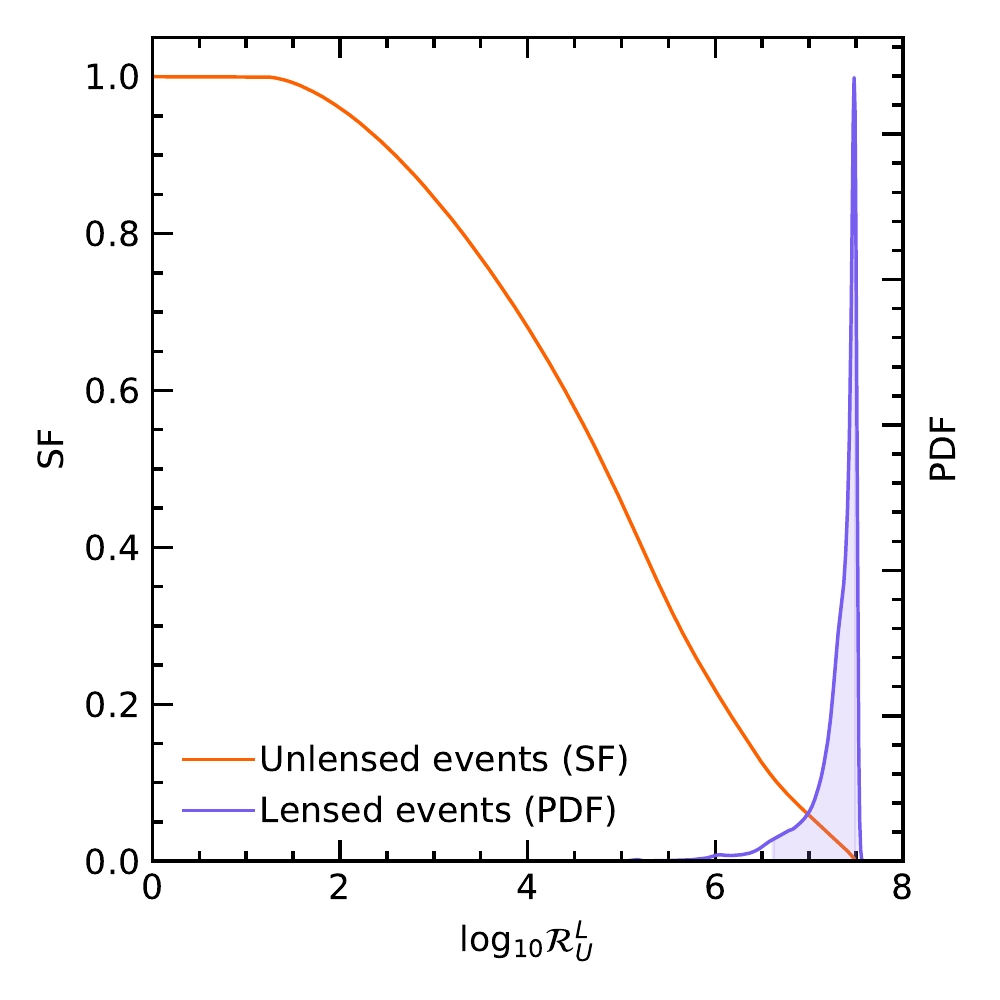}
    \caption{The $\RLU$ distributions of simulated unlensed (orange) and lensed (purple) event pairs. For unlensed event pairs, the survival function (SF) is shown, which is $1 - \text{CDF}$ (cumulative distribution function). Only a small fraction of unlensed events have an $\RLU$ similar or higher than lensed events. }
    \label{fig:RLU}
\end{figure}
\begin{equation}
    \RLU = \frac{p(\Delta t | \text{Lensed})}{p(\Delta t | \text{Unlensed})}\,,
\end{equation}
which quantifies how much more likely, \emph{a priori}, a certain arrival time difference between event pairs is under the lensed hypothesis than under the unlensed one. 
The time-delay $\Delta t$ can, in principle, refer to the expected time-delay between any permutation of the image combinations from a single event. Time delays from triple- or quadruple-image systems are expected to be correlated, and including these correlations would further improve the discriminatory power of strong lensing searches. However, we will neglect the correlations between time delays in the following, as we only aim to demonstrate the basic principle here.

As a practical example, we take the time-delay distribution to be for the difference in arrival time between any two consecutive images from quadruply lensed systems (Fig.~\ref{fig:time_delays}, right panel, gray shaded region). This equates to the hypothesis that two triggers come from a quadruply lensed event, but it is unknown where they place in the chronological order.

Let us first inspect the improvement in the significance of lensed detections due to the inclusion of the lensed time-delay prior. 
We simulate unlensed and lensed populations of events and compute the $\RLU$ for all event pairs. 
Based on the survival function (Fig.~\ref{fig:RLU}), we find a decrease of a factor of $3.1 \times 10^{-2}$, on average, in the false alarm probability per event pair produced by a randomly chosen lensed event, due to the inclusion of lensing time-delay information. 
That is, by incorporating the expected lensing time-delay distribution, the significance of lensed detections has improved, on average, by a factor of $\sim 32$. 

However, the benefit of incorporating the time-delay distribution becomes even more apparent when inspecting a catalogue of events. 
The total catalogue false alarm probability (the probability of finding at least one false alarm in a set of $N_{\text{pairs}}$ signal pairs) 
\begin{equation}
    \label{eqn:FAP}
    \FAP = 1 - \prod_{i = 0}^{N_{\text{pairs}}} (1 - p_i),
\end{equation}
where the false alarm per given event pair $p_i$ consists of an "intrinsic" false alarm probability, the probability that two events share a similar frequency evolution and thus mimic lensing by chance, and the probability that a lensed event produces a similar time-delay $\Delta t_i$ as the two unlensed events. 

\begin{figure}[!t]
    \centering
    \includegraphics[width = \linewidth]{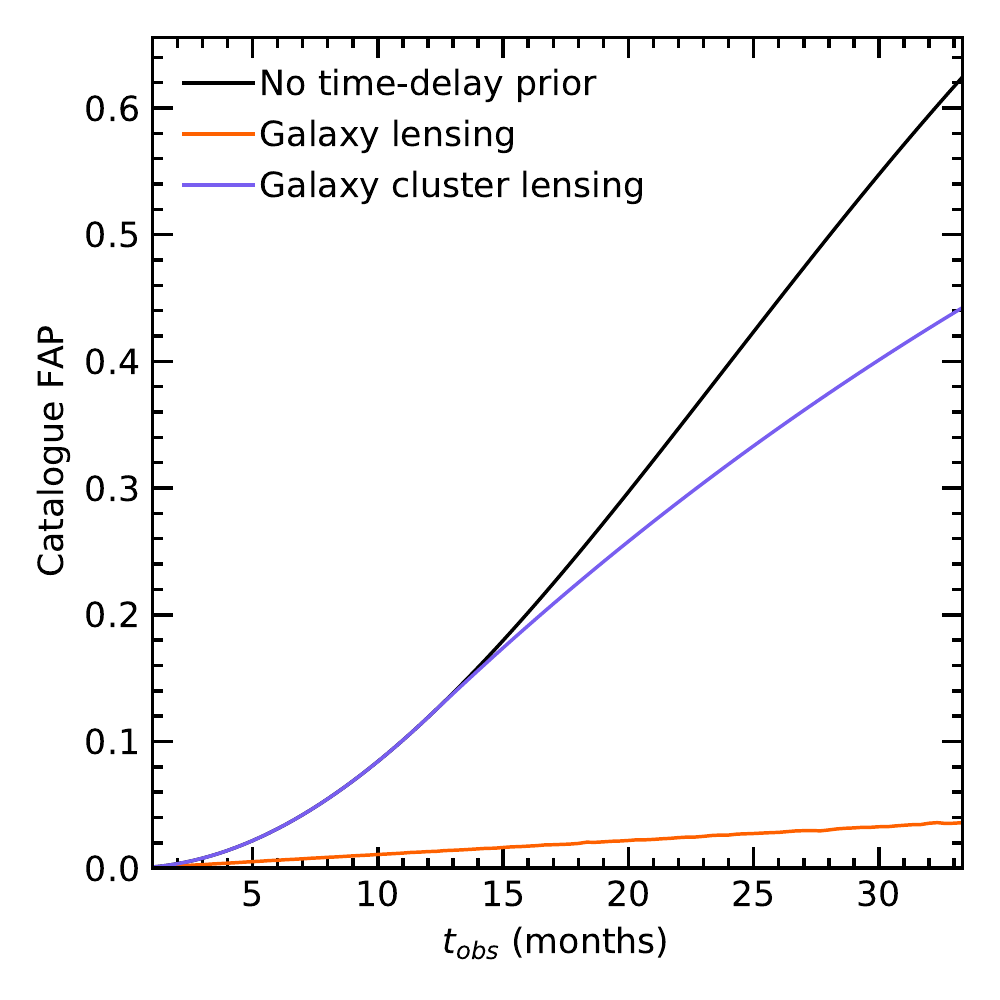}
    \caption{The catalogue false alarm probabilities (FAP) as a function of the observation run times $t_{obs}$, assuming a constant event rate $N = 510$ {events/yr}, time-delay distribution of all quadruples, time window $\Delta t_{\rm cluster} = 1$ {yr} and a FAP per event pair of $= 10^{-6}$. Shown are the FAP without windowing (black), the windowed FAP for galaxy cluster lensing (purple), and the ranked FAP for galaxy lensing (orange). Including galaxy lensing statistics changes the functional dependency in the exponential from $\propto t_{obs}^2$ to $\propto t_{obs}$. This reduces the FAP significantly for galaxy lensing when $t_{obs} \sim 1$ {yr}. For galaxy cluster lensing, the improvement is less significant owing to the longer lensing time delays. }
    \label{fig:fAP}
\end{figure}

Without incorporating knowledge of the lensing time delays, all events $N$ from the observing run need to be taken into account with equal weight, giving $N_{\text{pairs}} = N(N - 1)/2$, where $N$ is the total number of single events. 
This makes the likelihood of finding a false alarm inevitable as we obtain more gravitational wave detections (Fig.~\ref{fig:fAP}, black line).

\begin{figure}[!t]
    \centering
    \includegraphics[width = \linewidth]{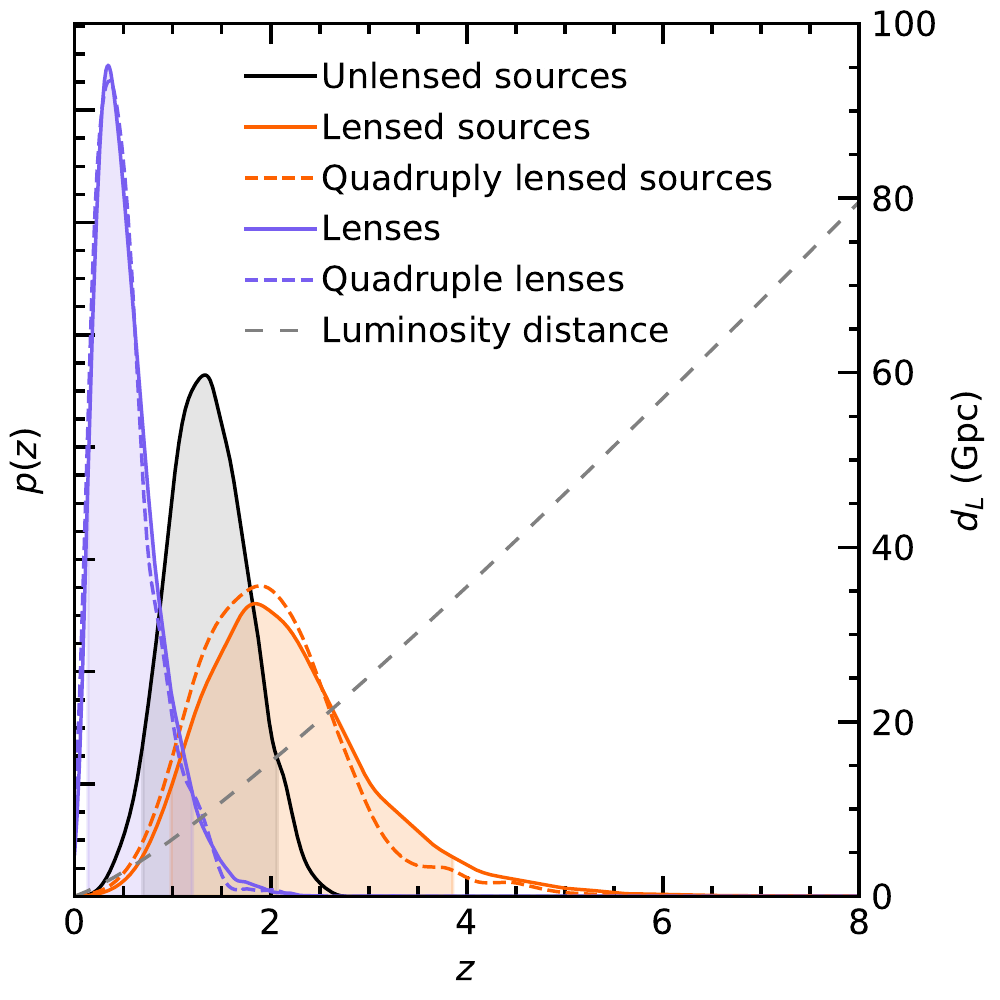}
    \caption{The observed redshift distributions for the galaxy lenses (purple) and lensed sources (orange), plotted on different scales. The unlensed source distribution (black) is shown for comparison. 
    Additionally, we show the distributions specifically for events that have been quadruply lensed (dashed), as an example of the versatility of the data. 
    The 90\% confidence interval for the unlensed sources is $z_s \sim 0.7 - 2.1$, while for the lensed sources $z_s \sim 1.0 - 3.9$. 
    Lensed events can thus allow us to probe events beyond the regular detector horizon.}
    \label{fig:redshift_distributions}
\end{figure}

However, when including galaxy lensing statistics, we find that the catalogue false alarm probability increases linearly with time, similar to typical single-event false alarms (Fig.~\ref{fig:fAP}, orange line). 
Indeed, we argue that prior knowledge of the lensing time delays not only offers an advantage in the strong lensing searches, but that it is necessary to enable the searches. 
Without prior knowledge of the time delays, the searches will inevitably run into false alarms.

The implications are particularly important when considering events with large time delays, such as the GW170104--GW170814 event pair investigated in~\citet{Dai:2020tpj,Liu:2020par,Abbott:2021iab}. 
Such events, if lensed, would be lensed by galaxy clusters, for which the lensing time-delay distribution is less well understood and the probability of a false alarm is significantly higher (Fig.~\ref{fig:fAP}, purple line). 
Here we assume a simple uniform prior between $0$ and $1 \, \rm yr$ for the time-delay of galaxy clusters, mostly for illustrative purposes. 
Therefore, we should be particularly careful in understanding the time-delay distribution and interpreting the results in light of the entire gravitational-wave catalogue for such events. 

Unfortunately, the time-delay distribution is subject to astrophysical uncertainties in lens modeling and the modeling of the binary population. 
Thus, we argue that careful follow-up investigations to understand the astrophysical uncertainties in modeling the statistical distribution of lensed events are vital to strong lensing searches. 
Detailed investigation of the false alarm probability in gravitational-wave catalogues will be given in (\c{C}al{\i}\c{s}kan et al., in preparation). 

Finally, we note that the inclusion of expected image types~\citep{Dai:2017huk} and relative magnifications~\citep{Lo:2021nae} may also improve the discriminatory power of strong lensing searches.
In our simulation, quadruple images typically consist of two subsequent type-I and two subsequent type-II images; the second and third images are type-I and type-II. 

\section{Redshift and lens distribution}
\label{sec:redshift}

\begin{figure}[!t]
    \centering
    \includegraphics[width = \linewidth]{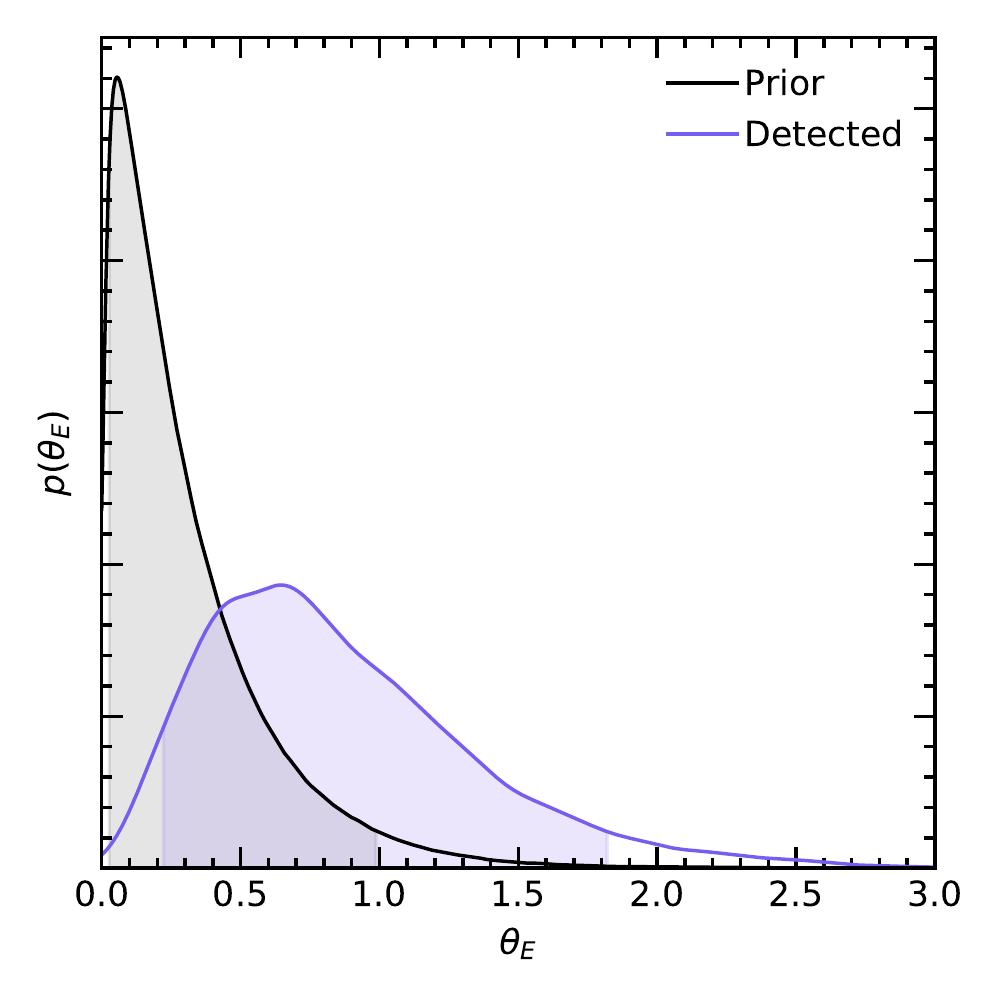}
    \caption{The distributions for Einstein radii of detected lensed events (purple) and the underlying population (black). The 90\% confidence interval for the detected Einstein radii is $\theta_E \sim 0.2 - 1.8$ arcsec, while the prior population generally has Einstein radii $\theta_E < 1.0$ arcsec. }
    \label{fig:einsteinradii}
\end{figure}

Strongly lensed gravitational waves originate from higher redshifts than unlensed gravitational waves. Particularly, lensed events originate from redshifts $z_s \sim 1.0 - 3.9$, above the usual detector horizon (Fig.~\ref{fig:redshift_distributions}). 
We note that strongly lensed gravitational-wave events can, in principle, be localised by combining gravitational-wave and electromagnetic measurements~\cite[e.g.,][]{Hannuksela:2020xor}. 
Thus, when localised, they may allow for high-redshift luminosity distance measurements. 
This may be particularly interesting for cosmology, where it has been suggested that some of the existing high-redshift luminosity distance measurements could be at odds with the standard $\rm \Lambda$CDM model~\cite[e.g.,][]{Risaliti:2018reu,Wong:2019kwg,DiValentino:2021izs}. 
However, we note that the localisation itself depends on the redshift distribution and the lens properties; only some fraction of host galaxies can be located in electromagnetic lensing surveys if they are near enough and their Einstein radii are large enough to be resolvable. 
We show the distribution of Einstein radii in Fig.~\ref{fig:einsteinradii}, which may be informative for such localisation studies. 
Besides the fundamental interest, the characterization of the lensed events is important in understanding the lensing science case.

\section{Conclusions}
\label{sec:conclusions}

Here we have reported 1) the expected number of double, triple, and quadruple gravitational-wave image detections in upcoming observing runs, 2) the positive impact of incorporating the lensing time-delay distribution on the false alarm probability for multi-image searches, 3) the expected source redshift and Einstein radius distribution of lensed gravitational-wave events. We have also demonstrated how using a galaxy (or galaxy cluster) lensing time-delay prior in our searches allows us to reduce the complexity of double-image searches. By including a prior, the false alarm probability increases linearly with time (similar to non-lensed searches) rather than exhibiting quadratic growth with time. 
However, more work is needed in modeling the merger-rate density, which is largely observationally unconstrained, in studying the precise improvement in the detection rates from sub-threshold searches and understanding the lensing time-delay distribution of events lensed by galaxy clusters. 

A lot of work on the forecasts has now been done and, besides our work, many groups have found reasonable rates of gravitational-wave lensing at the design sensitivity and beyond~\citep{Ng:2017yiu,Li:2018prc,Oguri:2018muv,Xu:2021bfn,Mukherjee:2021}. 
Further progress in estimating the precise rate will likely be impeded by the lack of binary black hole observations at high redshifts, where lensed gravitational waves originate from, although studies of the stochastic gravitational-wave background seem like a promising avenue~\citep{Buscicchio:2020bdq,Mukherjee:2020tvr,Buscicchio:2020cij,Abbott:2021iab}. 
Nevertheless, we expect that direct gravitational-wave lensing observations will give the final verdict on the rates. 
In the meantime, statistical forecasts can inform us of our tentative expectations, allow us to efficiently investigate the science case and potential improvements in search methodologies, and offer mock data simulations to stress-test our tools.
To facilitate such follow-up research, we have published our catalogue of simulated lensed gravitational-wave events in~\citet{wierda_a_renske_a_c_2021_4905030}. 
We also hope that our work gives further motivation to include lensing statistics results in strong lensing searches. \\

\acknowledgements
\noindent
\textbf{Acknowledgements} 
The authors thank Alvin Li, Chun-Lung Chan, Manchun Yeung, and Tjonnie Li for useful comments and feedback. 
The authors also thank Jose Ezquiaga for detailed comments on the manuscript. 
We also thank Thomas E. Collett, Mesut \c{C}al{\i}\c{s}kan, Daniel Holz, Anupreeta More, Haris K, Riccardo Buscicchio, and Jolien Creighton for discussion on related projects. 
A.R.A.C.W., O.A.H, and C.V.D.B. are supported by the research program of the Netherlands Organisation for Scientific Research (NWO).
The authors are grateful for computational resources provided by the LIGO Laboratory and supported by the National Science Foundation Grants No. PHY-0757058 and No. PHY-0823459.

\appendix
\onecolumngrid

\section{Derivation of the non-lensed and lensed rates} \label{app:lensed_rate_derivation}

\subsection{Non-lensed event rate}

The number of expected non-lensed gravitational-wave events per year can be expressed as an integral over the comoving volume
\begin{align}
    \label{eq:step1}
    \frac{\der N}{\der t} &= \int \frac{\der^2 N}{\der V_c \der t} 
    \frac{\der V_c}{\der z_s} \der z_s\,,
\end{align}
where $\der^2 N/(\der V_c \der t)$ is the merger-rate density measured in the detector frame, $\der V_c/\der z_s$ is the differential comoving volume, and $z_s$ is the redshift of the source binary black hole merger. 
The output of theoretical predictions and observational papers is the merger-rate density measured in the source frame $\mathcal{R}(z_s)=\der^2 N/(\der V_c \der t_s)=(1+z_s) [ \der^2 N/(\der V_c \der t)]$. 
Therefore, we express the integral in terms of the merger-rate density in the source frame
\begin{align}
    \frac{\der N}{\der t} &= \int \frac{\mathcal{R}(z_s)}{1+z_s}
    \frac{\der V_c}{\der z_s} \der z_s\,.
\end{align}

On the other hand, not all mergers are observed. 
Instead, only a fraction of signals at redshift $z_s$ with a network signal-to-noise ratio (SNR) larger than a detection network SNR threshold $\rho_c$ are observed 
\begin{equation}
    P(\rho>\rho_c|z_s) = \int_0^\infty \Theta(\rho(z_s,\vtheta) - \rho_c) p(\vtheta) d\vtheta\,,
\end{equation}
where $\rho(z_s,\vtheta)$ is the network SNR of a signal with some binary parameters $\vtheta$,  $\Theta(\rho(z_s,\vtheta)>\rho_c)$ is the Heaviside step function, and $p(\vtheta)$ is the expected distribution of binary parameters. 
Therefore, the rate of \emph{observed} mergers is 
\begin{align} \label{eq:non_lensed_rate}
    \frac{\der N_{\rm obs}}{\der t} &= \int \frac{\mathcal{R}(z_s)}{1+z_s}  \Theta(\rho(z_s,\vtheta) - \rho_c)
    \frac{dV_c}{dz_s} p(\vtheta) \der\vtheta \der z_s\,.
\end{align}
We adopt the \textsc{IMRPhenomD} waveform with aligned spins~\citep{Husa:2015iqa, Khan:2015jqa} in the network SNR computation. 
For the standard procedure to compute the network SNR, see, e.g.,~\citet{Roulet_2020}.

\subsection{Lensed event rate}

The lensed event rate follows the same idea, except that 
1) only a fraction of of gravitational waves are lensed, and 
2) the events can be multiply imaged and magnified. This essentially translates to a change of the merger-rate density in Eq. \eqref{eq:step1}
\begin{equation}
    \frac{\der^2 N}{\der V_c \der t} \to \frac{\der^2 N_{\rm obs}^{\rm SL}}{\der V_c \der t} = \frac{\der N_{\rm obs}^{\rm SL}}{\der N} \frac{\der^2 N}{\der V_c \der t}
\end{equation}
where $\der N_{\rm obs}^{\rm SL}/\der N$ is the fraction of observed lensed events with respect to the total events $N$. It is composed of the probability that a source at redshift $z_s$ is lensed times the fraction of detected images
\begin{equation}
    \label{eq:lensed_fraction}
    \frac{\der N_{\rm obs}^{\rm SL}}{\der N} 
    = \int \left (\sum_i^{\text{images}} \Theta(\rho(z_s, \vtheta, \mu_i, \Delta t_i) - \rho_c) \right) \times p(\text{SL}, \vtheta_L, z_L, \vbeta | z_s) p(\vtheta) \der \vtheta \der \vtheta_L \der z_L \der \vbeta \,,
\end{equation}
with $\mu_i$ and $\Delta t_i$ the $i$-th magnification and time-delay of a source at redshift $z_s$ due to a lens at redshift $z_L$, with lens parameters $\vtheta_L$ and source position in the lens plane $\vbeta$. 
The sum enforces detectability of the individual images, while $p(\text{SL}, \vtheta_L, z_L, \vbeta | z_s)$ represents the fraction of lenses at redshift $z_L$ with parameters $\theta_L$ that strongly lens a source at a given redshift $z_s > z_L$  for a source position in the lens plane $\vbeta$. We further break the probabilities in Eq. \eqref{eq:lensed_fraction} as follows
\begin{equation}
    p(\text{SL}, \vtheta_L, z_L, \vbeta | z_s) = \tau(z_s) p(\vtheta_L, z_L, \vbeta | \text{SL}, z_s) \,,
\end{equation}
where we introduced the optical depth $\tau(z_s) = p(\text{SL} | z_s)$.
We assume $\vbeta$ to be independent of $z_s$, $z_L$ and $\vtheta_L$, allowing us to write $p(\vtheta_L, z_L, \vbeta | \text{SL}, z_s) = p(\vtheta_L, z_L | \text{SL}, z_s) p(\vbeta | \text{SL})$.
Altogether, this gives us the observed lensed trigger rate in terms of the source frame merger rate density
\begin{equation}
    \label{eq:lensed_rate}
    R_{\rm SL} = \int \frac{\mathcal{R}(z_s)}{1+z_s} \tau(z_s) \left( \sum_i^{\text{images}} \Theta(\rho_i(\vtheta,z_s, z_L, \vtheta_L, \beta) - \rho_c) p(\vtheta)\right) \times p(\vtheta_L, z_L | \text{SL}, z_s) p(\vbeta | \text{SL}) \frac{\der V_c}{\der z_s} \der \vtheta \der \vbeta \der z_l \der \vtheta_L \der z_s\,.
\end{equation}

Note that in Tables~\ref{tab:rates} and~\ref{tab:subrates} we quote the number of detectable \emph{events}, and not the number of detectable images. We require at least two images to pass the SNR threshold for an event to be detectable, and count detectable events only once in the sum, as opposed to having all its detectable images add to the sum.

\subsection{Solving the rates integral}

We use Monte-Carlo integration with importance sampling to solve the integral in Eq. \eqref{eq:lensed_rate}. This method is based on the principle that
\begin{equation}
    \int f(x) p(x) dx \approx \frac{1}{N} \sum_{x_i \text{ from } p(x)} f(x_i) \,,
\end{equation}
so that solving the integral can be done by sampling from the respective probability distributions. We will use this approach to sample all of the parameters in Eq. \eqref{eq:lensed_rate} for one million systems. Effectively, this means we will create a population of $10^6$ binary black holes, and assign lenses to each of them to create a strong lensing configuration. We will explain these steps in Appendices \ref{app:bbh} and \ref{app:lenses} respectively.

\section{Assembling the binary black hole population}
\label{app:bbh}

The parameters that define a binary black hole merger are: source frame masses $m_1$ and $m_2$, orbital plane inclination $\iota$ and polarisation $\psi$, redshift $z_s$, sky localisation $\ra$ and $\dec$ and the arrival time $t$. The sky localisation is uniformly distributed across the celestial sphere, and the arrival time uniformly throughout the span of 1 yr. The polarisation follows a uniform distribution between 0 and $2\pi$, while the inclination is sampled from $p(\iota) = 0.5\sin(\iota)$ on the domain $[0, \pi]$.

\subsection{Sampling the mass distribution}
\label{sub:m1}

Sampling the source frame masses is a less trivial exercise. An inference of the true mass distribution is done in \citet{GWTC2:rates} with the events from GWTC-2. They investigated four different mass models, but we will only use the \textsc{Power-law + Peak} model for our research. This model is motivated by the possibility of a pile-up before the pair-instability gap, due to the mass loss in pulsational pair-instability supernovae.

The probability distribution is broken down according to $p(m_1, q|\theta_{\rm pop}) = p(q|m_1, \theta_{\rm pop}) p(m_1|\theta_{\rm pop})$, with $q$ the mass ratio $m_2/m_1$ and $\theta_{\rm pop}$ the underlying population parameters. The distribution for $m_1$ is given by
\begin{equation}
    \label{eqn:m1pdf}
    p(m_1|\lambda_{\rm peak}, \alpha, m_{\rm max}, m_{\rm min}, \delta_m, \mu_m, \sigma_m) = \left[ (1 - \lambda_{\rm peak}) \mathfrak{P}(m_1|-\alpha, m_{\rm max}) + \lambda_{\rm peak} G(m_1|\mu_m, \sigma_m)\right] S(m_1|m_{\rm min}, \delta_m) \,,
\end{equation}
with $\mathfrak{P}$ a normalised power-law distribution with spectral index $-\alpha$ and cut-off $m_{\rm max}$. $G$ is a Gaussian distribution with mean $\mu_m$ and width $\sigma_m$, and the parameter $\lambda_{\rm peak}$ gives the fraction of binaries that follow the Gaussian. Finally, $S$ is a smoothing function, which is defined as
\begin{align*}
    S(m|m_{\rm min}, \delta_m) = & 
    \begin{cases}
    0 & m < m_{\rm min} \\
    [f(m - m_{\rm min}, \delta_m) + 1]^{-1} & m_{\rm min} \leq m < m_{\rm min} + \delta_m \\
    1 & m \geq m_{\rm min} + \delta_m
    \end{cases} \\
    f(m', \delta_m) = & \exp\left(\frac{\delta_m}{m'} + \frac{\delta_m}{m' - \delta_m} \right) \,.
\end{align*}

\twocolumngrid

The distribution for the mass ratio is defined for $q \leq 1$ and is given by
\begin{equation}
    p(q|\beta, m_1, m_{\rm min}, \delta_m) \propto q^\beta S(qm_1|m_{\rm min}, \delta_m) \,,
\end{equation}
with $\beta$ the spectral index of the power-law. Because of the complex nature of this combined distribution, we will break down the sampling step-by-step.
\begin{figure}[!b]
    \includegraphics[width = \linewidth]{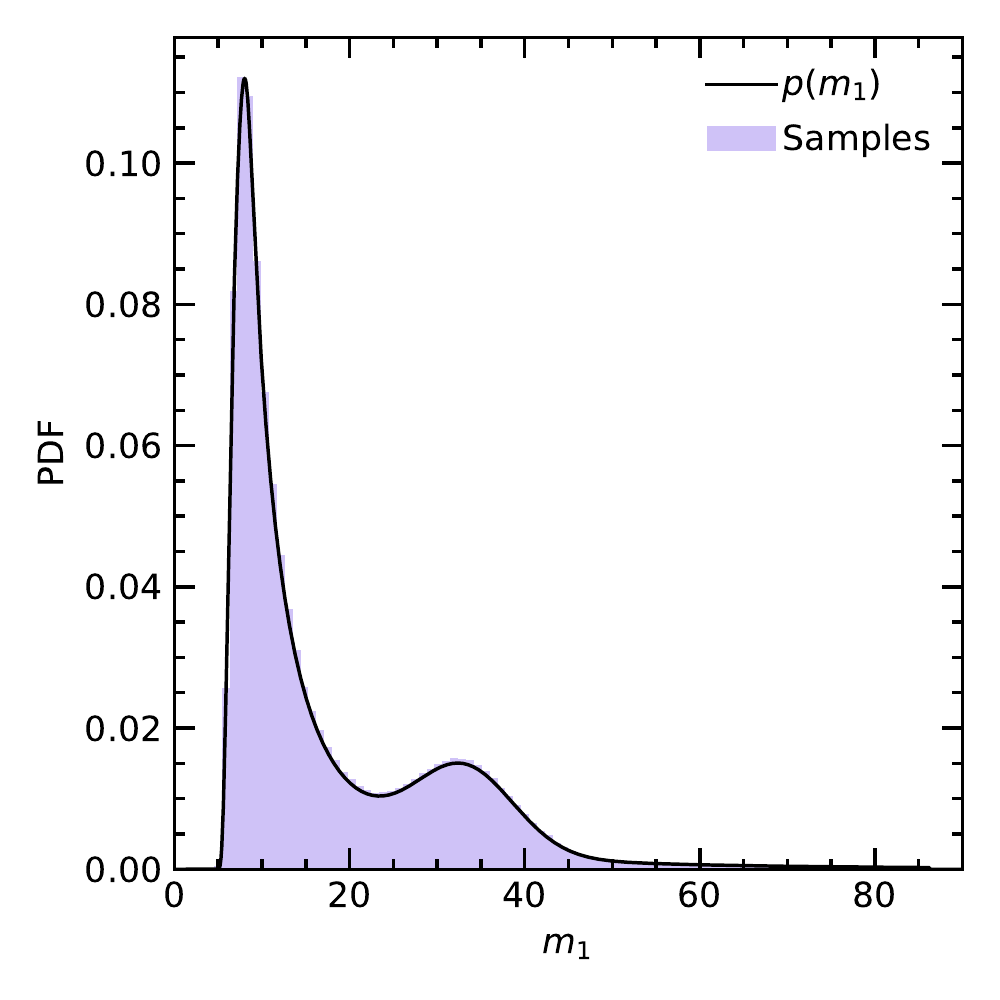}
    \caption{An illustration of the \textsc{Power-law+peak} prior probability distribution for the primary mass $m_1$ for fixed values of the hyper-parameters.}
    \label{fig:m1pdf}
\end{figure}

We can break down Eq. \eqref{eqn:m1pdf} as follows: the total population consists two sub-populations, that follow their respective distributions $\mathfrak{P}$ and $G$. Before sampling these, we first draw a random number between 0 and 1 to determine which fraction of the total population gets sampled. If it is smaller than $\lambda_{\rm peak}$, then $G$ gets sampled, and vice-versa. This will give us values for $m_1$, but the total distribution still needs to be tailored by the smoothing function $S$ at the low-mass end of the spectrum. We fix this by drawing random numbers $u$ between 0 and 1, and rejecting samples where $u > S(m_1, m_{\rm min}, \delta_m)$. This will get rid of any excess low-mass samples. This whole procedure gives us a sample set that follows the correct pdf, as we can see in Fig.~\ref{fig:m1pdf}. Sampling $m_2$ follows a very similar structure, with the simplification that there are no sub-populations. 

\begin{table}[!h]
    \centering
    \begin{tabular}{r  l}
        \hline
        Parameter & Value \\ 
        \hline \hline
        $\lambda_{\rm peak}$ & 0.10 \\
        $\alpha$ & 2.63 \\
        $\beta$ & 1.26 \\
        $\mu_m$ & 33.07 $M_\odot$ \\
        $\sigma_m$ & 5.69 $M_\odot$ \\
        $m_{\rm max}$ & 86.22 $M_\odot$ \\
        $m_{\rm min}$ & 4.59 $M_\odot$ \\
        $\delta_m$ & 4.82 $M_\odot$ \\
        \hline
    \end{tabular}
    \caption{All the values of the model parameters for the BBH mass distributions, based on the results from the GWTC-2 population results~\citep{GWTC2:rates}.}
    \label{tab:m1params}
    \begin{tabular}{r l}
        \hline
        Parameter & Value \\
        \hline \hline
        $\lambda_z$ & 0.563 \\
        $a$ & 2.906 \\
        $b$ & 0.0158 \\
        $c$ & 0.58 \\
        $\mu_z$ & 1.1375 \\
        $\sigma_z$ & 0.8665 \\
        \hline
    \end{tabular}
    \caption{All the values of the parameters for the semi-analytical approximation to $\der V_c/\der z_s$.}
    \label{tab:zparams}
\end{table}

\subsection{Sampling the binary black hole redshifts}

The last binary black hole parameter we need to sample is the source redshift. We assume the binaries follow the differential comoving volume $dV_c/dz_s$, which can be normalised to give us $p(z_s)$. This normalisation is done on the domain $z_s \in [0, 10]$, as we do not expect any observable binaries outside of this region. We develop a semi-analytical approximation to $p(z_s)$ to accommodate for inverse transform sampling. This approximation is given by $p(z_s) \simeq\lambda_z f(z_s|a, b, c) + (1 - \lambda_z) g(z_s|\mu_z, \sigma_z)$, where $f$ is a beta prime distribution centred at $c$ and
\begin{equation}
    \label{eq:redshiftpdf}
    g(z) \propto \exp\left(-\frac{(\log(z) - \mu_z)^2}{2\sigma_z^2}\right) \,.
\end{equation}
This approximation resembles the situation with Eq. \eqref{eqn:m1pdf}, with the absence of a smoothing function. We can thus draw a number between 0 and 1 to choose a distribution based on $\lambda_z$, and sample the two distributions individually. Note that these both have to be normalised on the same domain as $\der V_c/\der z_s$, because their formal domains are $z_s \in [0, \infty]$.

With this, we can now sample all necessary binary black hole parameters. We repeat this 1 million times, and giving us a catalogue of binary black hole mergers.

\section{Creating the lensed population}
\label{app:lenses}

The parameters that define a PEMD (Power-law Elliptical Mass Distribution) galaxy lens are: velocity dispersion $\sigma$, axis ratio $q$, axis rotation $\psi$, and spectral index of the density profile $\gamma$. We add to this an external shear, defined by $\gamma_1$ and $\gamma_2$, and place the galaxy at redshift $z_L$. Both shears are drawn from a normal distribution centred at 0 and with a width of 0.05~\citep{Collett:2015roa}. The axis rotation follows a uniform distribution between 0 and $2\pi$, while the density profile is sampled from a normal distribution with a width of 0.2, centred at 2~\citep{Koopmans:2009}. The sampling of the remaining parameters is (somewhat) dependent on the source redshift, so a source is picked from the previously compiled binary black hole catalogue.

The lens redshift is then sampled in multiple steps~\citep{Haris:2018vmn}. First, a value $r$ between 0 and 1 is drawn from the distribution
\begin{equation}
    p(x) = 30x^2(1 - x)^2 \,.
\end{equation}
The comoving distance to the lens $D^c_L$ is given by $D^c_L~\!=~\!rD^c_s$, with $D^c_s$ the comoving distance to the source. This can be translated to the redshift of the lens $z_L$. 

For the velocity dispersion, we sample a parameter $a$ from a generalised gamma distribution
\begin{equation}
    p(x) = x^{\alpha - 1}\exp(-x^\beta)\frac{\beta}{\Gamma(\alpha/\beta)} \,,
\end{equation}
where $\alpha = 2.32$ and $\beta = 2.67$, and we take $\sigma~\!=~\!a~\times~161~{\rm km \, s^{-1}}$ \citep{Collett:2015roa}. 
We use the individual lensing probability to condition our distributions on strong lensing, as is required in $p(\theta_L, z_L | \text{SL}, z_s)$. We calculate the Einstein radius $\theta_E$ through
\begin{equation}
    \theta_E = \frac{4\pi\sigma^2}{c^2}\frac{D_{Ls}}{D_s} \,,
\end{equation}
with $D_{Ls}$ and $D_s$ the angular diameter distances between lens and source and observer and source, respectively. The individual lensing probability is then given by $p(SL|\theta_L, z_L, z_s) = \pi \theta_E^2/4\pi \propto \theta_E^2$. All lenses are rejections sampled, where we draw a uniformly distributed number between 0 and $3^2$ Einstein radii and pass those that have a value~$< \theta_E^2$.

Finally, we draw a parameter $b$ from a Rayleigh distribution with scale\footnote{\citet{Collett:2015roa} has a typo in the scaling parameter, the original \textsc{LensPop} code has the correct scaling we assume here.} $s = 0.38 - 0.09177a$, 
\begin{equation}
    p(x) = \frac{x}{s^2} \exp \left(-\frac{x^2}{2s^2} \right) \,,
\end{equation}
which is sampled until we get a value $b < 0.8$. The axis ratio is then given by $q = 1 - b$, which concludes the sampling of the lens parameters.

We also need to draw a source position in the lens plane $\vbeta$, in order to solve the lens equation. Since we are only interested in strong lensing configurations, we draw uniformly distributed positions around/inside the lens area until we get a solution with 2 or more images. This effectively incorporates $p(\vbeta|SL)$ from Eq. \eqref{eq:lensed_rate}.
We compute the image time delays and magnifications for each sample using \textsc{lenstronomy}~\citep{Birrer:2018xgm}.

We repeat this whole process for a million randomly chosen sources from the binary black hole catalogue, and save the results in a separate catalogue. Combining the two, we get our final lensed catalogue. All events are assigned a weight $\tau(z_s) \mathcal{R}(z_s)/(1 + z_s)$, which gives their true relative occurrence, and is used to quantify the importance of each event.

\bibliographystyle{aasjournal}

\end{document}